\documentclass{article}

\newcommand{\p}[1]{(\ref{#1})}

\newcommand{\be}{\begin{equation}}
\newcommand{\ee}{\end{equation}}
\newcommand{\bea}{\begin{eqnarray}}
\newcommand{\eea}{\end{eqnarray}}

\newcommand{\ba}{\begin{array}} \newcommand{\ea}{\end{array}}

\newcommand{\nn}{\nonumber}

\usepackage{amscd,amsmath,amssymb}

\def\theequation{\arabic{section}.\arabic{equation}}

\begin{document}
\thispagestyle{empty}
\vspace{2cm}
\begin{flushright}
%Version 6.0 \\
%\today \\
\end{flushright}\vspace{2cm}
\begin{center}
{\Large\bf Supersymmetric component actions via coset approach}
\end{center}
\vspace{1cm}

\begin{center}
{\large\bf S.~Bellucci${}^a$, S.~Krivonos${}^{b}$ and A.~Sutulin${}^{b}$}
\end{center}

\begin{center}
${}^a$ {\it
INFN-Laboratori Nazionali di Frascati,
Via E. Fermi 40, 00044 Frascati, Italy} \vspace{0.2cm}

${}^b$ {\it
Bogoliubov  Laboratory of Theoretical Physics, JINR,
141980 Dubna, Russia} \vspace{0.2cm}

\end{center}
\vspace{2cm}

\begin{abstract}\noindent
We propose a method  to construct the on-shell component actions for the theories with $1/2$ partial
breaking of global supersymmetry within the nonlinear realization (coset) approach. In contrast with the standard superfield approach in which unbroken supersymmetry plays the leading role,  we have shifted the attention to the spontaneously broken supersymmetry. It turns out that in the theories in which half of supersymmetries is spontaneously broken, all physical fermions are just the fermions of the nonlinear realization. Moreover, the transformation properties of these fermions with respect to the broken supersymmetry are the same as in the famous Volkov-Akulov model. Just this fact completely fixed all possible appearances of the fermions in the component action: they can enter the action  through the determinant of the vielbein (to compensate the transformation of the volume form) and the covariant derivatives, only. It is very important that in our parametrization of  the coset the rest of physical components, i.e. all  bosonic components, transform as  ``matter fields'' with respect to the broken supersymmetry. Clearly, in  such a situation the component action acquires the form of the Volkov-Akulov action for these ``matter fields''. The complete form of the action can be further fixed by two additional requirements: a) to reproduce the bosonic limit, which is explicitly known in many interesting cases, and b) to have a proper linearized form, which has to be invariant with respect to the linearized unbroken supersymmetry.
We supply the general consideration by a detailed example of the component action of $N=1$ supermembrane in $D=4$ constructed within our  procedure. In this case we provide  the  exact proof of the invariance of the constructed component action with respect to both, broken and unbroken supersymmetries.
\end{abstract}

\newpage
\setcounter{page}{1}
\setcounter{equation}{0}
\section{Introduction and general setup}
It is a well known fact that a domain wall spontaneously breaks the Poincar\'{e} invariance of the target space
down to the symmetry group of the world volume subspace. This breaking results in the appearing of the Goldstone bosons
associated with spontaneously broken symmetries. When we are dealing with the purely bosonic $p$-branes this information is enough to construct the corresponding action\footnote{The situation with the bosonic $D$-branes, which necessarily contain the gauge fields is less clear, despite the knowledge of the explicit actions, etc \cite{dbrane}. Fortunately, in the supersymmetric cases, where the supersymmetry is also partially spontaneously broken, the bosonic sector, which is the combination of Nambu-Goto and Born-Infeld actions, appears automatically.}. From the mathematical point of view, the most appropriate approach to describe a partial breaking of Poincar\'{e} symmetry is the nonlinear realization (or coset)
method \cite{NR} suitably modified for the cases of (supersymmetric) space-time symmetries in \cite{NR1}. Schematically the coset approach works as follows.

After splitting the generators of the target space $D$-dimensional Poincar\'{e} group, which is supposed to be spontaneously broken on the world volume  down to the $d$-dimensional Poincar\'{e} subgroup, into the generators of unbroken\footnote{For the sake of brevity we suppress here all space-time indices.}  $\{P , M\}$ and spontaneously broken   $\{Z, K\}$ symmetries (the generators $P$ and $Z$ form $D$-dimensional translations, $M$ generators span the
$so(1,d-1)$ - Lorentz algebra on the world volume, while generators $K$ belong to the coset $so(1,D-1)/so(1,d-1)$), one may realize all the transformations of $D$-dimensional Poincar\'{e} group by the left action on the
coset element
\be\label{I1}
g= e^{x P} e^{q(x) Z} e^{\Lambda(x)K}.
\ee
The spontaneous breaking of $Z$ and $K$ symmetries is reflected in the character of corresponding coset coordinates which
are Goldstone fields $q(x)$ and $\Lambda(x)$ in the present case. The transformation properties of  coordinates $x$ and
fields $q(x), \Lambda(x)$ may be easily found in this approach, while all information about geometric properties is contained in the Cartan forms
\be\label{I2}
g^{-1}d q = \Omega_P P + \Omega_M M+ \Omega_Z Z+\Omega_K K.
\ee
All Cartan forms except for $\Omega_M$ are transformed homogeneously under all symmetries.
Due to the general theorem \cite{ih} not all of the above Goldstone fields have to be treated as independent. In the present
case the fields $\Lambda(x)$ can be covariantly expressed through $x$-derivatives of $q(x)$ by imposing the constraint
\be\label{I3}
\Omega_Z=0.
\ee
Thus, we are dealing with the fields $q(x)$ only. It is very important that the form $\Omega_P$ defines the vielbein $E$ ($d$-bein  in the present case), connecting the covariant world volume coordinate differentials $\Omega_P$ and the world volume coordinate differential $dx$ as
\be\label{I4e}
\Omega_P = E \cdot dx .
\ee
Combining all these ingredients, one may immediately write the  action
\be\label{I4}
S = -\int d^dx + \int d^dx\; \det(E),
\ee
which is invariant under all  symmetries. In \p{I4} we have added the  trivial first term to fulfill  the condition $S_{q=0}=0$. The action \p{I4} is just the static gauge form of the actions of $p=(D-d)$-branes.

The supersymmetric generalization of the coset approach involves into the game new spinor generators $Q$ and $S$ which extend the
$D$-dimensional Poicar\'{e} group to the supersymmetric one
\be\label{I5}
\left\{ Q,Q\right\} \sim P, \; \left\{ S,S\right\} \sim P, \; \left\{ Q,S\right\} \sim Z.
\ee
The most interesting cases are those  when the $Q$ supersymmetry is kept unbroken, while the $S$ supersymmetry is supposed to be spontaneously broken\footnote{If all supersymmetries are considered as spontaneously broken, the corresponding action can be constructed similarly to the bosonic case, resulting in the some synthesis of Volkov-Akulov \cite{VA} and Nambu -Goto actions. An enlightening example of such a construction can be found in \cite{CNV}}. When $\#Q=\#S$ we are facing the so called $1/2$ Partial Breaking of Global Supersymmetry cases,  which most of all interesting supersymmetric domain walls belong to. Only such cases of supersymmetry breaking will be considered in this Letter.

Now, all our symmetries can be realized by group elements acting on the coset element
\be\label{I6}
g= e^{x P} e^{\theta Q} e^{q(x,\theta) Z} e^{\psi(x,\theta)S} e^{\Lambda(x,\theta)K}.
\ee
The main novel feature of the supersymmetric coset \p{I6} is the appearance of the Goldstone superfields $\{q(x,\theta), \psi(x,\theta), \Lambda(x,\theta)\}$ which depend on the coordinates of the world volume superspace $\{x, \theta\}$. The rest
of the coset approach machinery works in the same manner: one may construct the Cartan forms \p{I2} for the coset \p{I6}
(which will contain the new forms $\Omega_Q$ and $\Omega_S$), one may find the supersymmetric $d$-bein and corresponding
bosonic $\nabla_P$ and spinor $\nabla_Q$ covariant derivatives, etc. One may even  write the proper generalizations of the covariant constraints \p{I3} as
\be\label{I7}
\Omega_Z=0, \quad \Omega_S|=0,
\ee
where $|$ means the $d\theta$-projection of the form (see e.g. \cite{Iv} and references therein).
The $d\theta$ parts of these constraints are closely related with the "geometro-dynamical" constraint of the super-embedding approach (see e.g. \cite{Dima}).

Unfortunately, this similarity between purely bosonic and supersymmetric cases is not complete due to the existence of the following important new features of theories with partial breaking of global supersymmetry:
\begin{itemize}
\item In contrast with the bosonic case, not all of the physical fields appear among the parameters of the coset. A famous example comes from the supersymmetric $D3$-brane (aka $N=1$ Born-Infeld theory) where the coset element \p{I6} contains only $P,Q$ and $S$ generators \cite{BG2}, \cite{RT}, while the field strength is ``hidden'' inside
    the superfield $\psi: F\sim \nabla_Q \psi|$. Nevertheless, {\it it is true} that the {\it all physical bosonic components} can be found in the  quantity $\nabla_Q \psi|$.
\item The supersymmetric generalization \p{I7} of the bosonic kinematic constraints \p{I3} in most cases contains
not only kinematic conditions, but also dynamic superfield equations of motion. A prominent example again may be found
in \cite{BG2}. Moreover, in many  cases it is unknown how to split these constraints into kinematical and
dynamical ones.
\item But the most unpleasant feature of the supersymmetric cases is that the standard methods of nonlinear realizations
fail to construct the superfield action! The main reason for this is simple: all that we have at hands are the covariant Cartan forms,  which we can construct the  superfield invariants from, while the superspace Lagrangian is not invariant. Instead it is shifted by the full spinor derivatives under unbroken and/or broken supersymmetries.
\end{itemize}
Therefore, all that we can do until now, within the supersymmetric coset approach, is
\begin{itemize}
\item to find the transformation properties of the superfields and  construct the covariant derivatives
\item to find the superfield equations of motion and/or covariant variants of irreducibility constraints.
\end{itemize}
That is why during recent years some new methods to construct the actions (in terms of superfield or  in terms of
physical components) have been proposed. Among them one should mention  the construction of the linear realization of
partially broken supersymmetry \cite{BG2}, \cite{RT}, \cite{IK1}, \cite{BIK1} and reduction from higher dimensional
supersymmetric $D$-brane action \cite{dbrane} to lower dimensions \cite{RK}.

In this Letter we will make one further step in the application of the supersymmetric coset approach, by demonstrating how on-shell component actions can be constructed within it. This construction is so simple that it can be schematically formulated just here.

The main idea is to start  with the Ansatz for the action  manifestly invariant with respect to {\it spontaneously broken supersymmetry}. Funny enough, it is rather easy to do, due to the following properties:
\begin{itemize}
\item in our parametrization of the coset element \p{I6} the superspace coordinates $\theta$ do not transform under broken supersymmetry. Thus, all components of superfields transform {\it independently},
\item the covariant derivatives $\nabla_P$ and $\nabla_Q$ are invariant under broken supersymmetry. Therefore, the bosonic physical components which are contained in $\nabla_Q \psi| $ can be treated as  ``matter fields''
    (together with the field $q(x,\theta)|$ itself)
    with respect to broken supersymmetry,
\item all physical fermionic components are just $\theta=0$ projections of the superfield $\psi(x, \theta)$ and
these components transform as the fermions of the Volkov-Akulov model \cite{VA} with respect to broken supersymmetry.
\end{itemize}
The immediate consequence of these facts is the conclusion that the physical fermionic components can enter the component on-shell
action through the determinant of the $d$-bein $E$ constructed with the help of the Cartan form $\Omega_P$ in the limit $\theta=0$,
or through the space-time $\nabla_P$ derivatives of the ``matter fields'', only. Thus, the most general Ansatz for
the on-shell component action, which is invariant with respect to spontaneously broken supersymmetry, has the form
\be\label{I8}
S=\int d^d x - \int d^dx \det(E) {\cal F}( \nabla_Q \psi|, \nabla_P q|).
\ee
Note, that the arguments of the function ${\cal F}$ are the bosonic physical components $\nabla_Q \psi|$ without
any derivatives and/or covariant space-time derivatives of $q$ (which, by the way, are also contained in $\nabla_Q \psi|$).
The explicit form of the function ${\cal F}$ can be fixed by  two additional requirements
\begin{enumerate}
\item The action \p{I8} should have a proper bosonic limit, which is known in almost all interesting cases.
One should note, that this limit for the action \p{I8} is trivial
$$ S_{bos}=\int d^d x\left( 1 -{\cal F}( \nabla_Q \psi|, \partial_P q)\right).$$
\item The action \p{I8} in the linear limit should possess a linear version of unbroken supersymmetry, i.e. it should be just a sum of the kinetic terms for all bosonic and fermionic components with the relative coefficients fixed by
    unbroken supersymmetry.
\end{enumerate}
These conditions completely fix the component action. Of course, as the final step, the invariance
with respect to unbroken supersymmetry has to be checked.

In the next Section we apply the above procedure to the simplest case of $N=1$ supermembrane in $D=4$. In  Section 3, for completeness, we extend our analysis to the  case of the dual system - N=1 supersymmetric space filling $D2$ brane. Two Appendices contain the technical details, notation and explicit proof of the invariance of the supermembrane action with respect to both, broken and unbroken supersymmetries. We conclude with some comments and perspectives.
\setcounter{equation}{0}
\section{Supermembrane in D=4}
As the first instructive example of our approach we will consider in this Section the supermembrane
in $D=4$. In the first  subsection we will mainly follow  the paper \cite{IK1}.

\subsection{Kinematical constraints and equations of motion}
The nonlinear realization of the breaking $N=1, D=4 \rightarrow N=1,d=3$ has been constructed in \cite{IK1}.
There, the $N=1, D=4$ super Poincar\'{e} group has been realized in its coset over the $d=3$ Lorentz group $SO(1,2)$
\be\label{cosetA}
g=e^{x^{ab}P_{ab}}e^{\theta^{a}Q_{a}}e^{qZ}
  e^{\psi^aS_a}e^{\Lambda^{ab}K_{ab}} \;.
\ee
Here, $x^{ab}, \theta^a$ are $N=1, d=3$ superspace coordinates, while the remaining coset parameters are Goldstone superfields, $\psi^a \equiv \psi^a(x,\theta),\;q \equiv q(x,\theta),\; \Lambda^{ab}
\equiv \Lambda^{ab}(x,\theta)$.
To reduce the number of independent superfields one has to impose the constraints\footnote{We collect the exact expressions for the covariant derivatives $\nabla_{ab}, \nabla_a$ and their properties, constructed in \cite{IK1}, in  Appendix A.}
\be
\Omega_Z = 0 \quad \Rightarrow \quad \left\{
\begin{array}{l}
\nabla_{ab}q+ \frac{4}{1+2\lambda^2}\lambda_{ab} = 0 \quad\; \mbox{(a)}\\
\nabla_a q - \psi_a = 0\; \qquad \qquad \mbox{(b)}
\end{array}\right. \label{ihA}
\ee
The Eqs.\p{ihA} allow to express $\lambda_{ab}(x,\theta)$ and $\psi^a(x,\theta)$ through covariant derivatives of $q(x,\theta)$. Thus, the bosonic superfield $q(x,\theta)$ is the only essential Goldstone superfield we need for this case of the partial breaking of the global supersymmetry.
The constraints \p{ihA} are covariant under all symmetries and they do not imply any dynamics and leave $q(x,\theta)$ off shell.

The last step we can make within the coset approach is to write the covariant superfield equations of motion.
It was  shown in \cite{IK1} that this can be achieved  by imposing the following constraint on the Cartan form:
\be\label{eom_form}
\Omega_S| =0 \quad \Rightarrow \quad
\mbox{(a)}\;\; \nabla^a\psi_a = 0~, \quad \mbox{(b)}\;\;
\nabla_{(a}\psi_{b)}= -2\lambda_{ab}~.
\ee
where $|$ denotes the ordinary $d\theta$- projection of the form $\Omega_S$.

The equations \p{eom_form} imply the proper dynamical equation of motion
\be\label{peom}
\nabla^a \nabla_a q=0.
\ee
This equation is also covariant with respect to all symmetries, and its bosonic limit (for $q(x) \equiv
q(x,\theta)|_{\theta = 0}$) reads
\be  \label{NGeq}
\partial_{ab}\left( \frac{\partial^{ab}q}
       {\sqrt{1-\frac{1}{2}\partial q \cdot \partial q}}
   \right) =0~,
\ee
which corresponds to the ``static gauge'' form of the $D=4$
membrane Nambu-Goto action
\be  \label{NG}
S= \int d^3x \left( 1 - \sqrt{1-\frac{1}{2}\partial^{ab} q  \partial_{ab} q}
\right)\; .
\ee
Thus, the equations \p{eom_form} indeed describe the supermembrane in $D=4$.
\subsection{Component action for $N=1, D=4$ supermembrane}
Until now we just repeated the standard coset approach steps from the paper \cite{IK1} in the application to
the $N=1, D=4$ supermembrane. As we already said in the Introduction, the nonlinear realization approach  fails to construct the superfield action. That is why, to construct the superfield action one has to involve some additional arguments/scheme as it has been done, for example, in \cite{IK1}.

Funny enough, if we instead will be interested in the component action, then it can be constructed almost immediately within the nonlinear realization approach. One may check that all important features  of the on-shell (i.e. with the equations \p{eom_form} taken into account) component action we summarize in the Introduction, are present in the case at hands.
Indeed,
\begin{itemize}
\item all physical components, i.e. $q|_{\theta=0}$ and $\psi^a|_{\theta=0}$, are among the ``coordinates'' of our coset \p{cosetA} as the $\theta=0$ parts of the corresponding superfields,
\item under spontaneously broken supersymmetry the superspace coordinates $\theta^a$ do not transform at all \p{susy2}.
Therefore, the corresponding transformation properties of the fermionic components $\psi^a|_{\theta=0}$ are {\it the same as in the Volkov-Akulov model} \cite{VA}, where  all supersymmetries are supposed to be spontaneously broken,
\item Finally, the $\theta=0$ component of our essential Goldstone superfield $q(x,\theta)$ does not transform
under spontaneously broken supersymmetry and, therefore, it behaves like a ``matter'' field within the Volkov-Akulov
scheme.
\end{itemize}
As the immediate consequences of these features we conclude that
\begin{itemize}
\item The  fermionic components $\psi^a|_{\theta=0}$ may enter the component action either through $\det(E)$ \p{E}  (to compensate the transformation of volume $d^3 x$ under \p{susy2}) or through the covariant derivatives $\nabla_{ab}$ \p{nabla}, only,
    \item The ``matter'' field -- $q|_{\theta=0}$ may enter the action only through covariant derivatives $\nabla_{ab} q$.
\end{itemize}

Thus, the unique candidate to be the component on-shell action, invariant with respect to spontaneously broken supersymmetry $(S_a)$ reads
\be\label{cact1}
S= \alpha \int d^3x +\beta \int d^3x \left( \det E \right) {\cal F}(\nabla^{ab} q \nabla_{ab}q),
\ee
with an arbitrary, for the time being, function $\cal F$. All other interactions between the bosonic component $q$ and the fermions of spontaneously broken supersymmetry $\psi^a$ are forbidden!

Note, that the first, trivial term in \p{cact1} is independently invariant under broken (and unbroken!) supersymmetries, because,
in virtue of \p{susy2}
\be
\delta_S \int d^3x \sim \int d^3x \; \partial_{ab} \left( \xi^a \psi^b\right)\quad \mbox{and, therefore }\quad
\delta_S \int d^3x=0.
\ee
As we already said in the Introduction, this  term in the action \p{cact1} ensures the validity of the  limit
$S_{q=0,\psi=0} =0$.

The action \p{cact1} is the most general component action invariant with respect to unbroken supersymmetry. But in the present case we explicitly know its bosonic limit - it should be just the Nambu-Goto action \p{NG}. Some additional
information about its structure comes from the linearized form of the action, which,  according with its invariance with respect to unbroken supersymmetry, has to be
\be
S_{lin}\sim \psi^a \partial_{ab} \psi^b - \frac{1}{4} \partial^{ab}q \partial_{ab}q.
\ee
Combining all these ingredients, which completely fix the parameters $\alpha$ and $\beta$ in \p{cact1}, we can write the component action of $N=1, D=4$ supermembrane as
\be\label{Action}
S= \int d^3 x\left[ \;2 - \det(E)\;\left(1+\sqrt{1-\frac{1}{2} \nabla^{ab} q \nabla_{ab} q} \right)\right].
\ee
The explicit expression for $\det(E)$ has the form
\bea\label{detE}
\det(E)&=& 1+\frac{1}{2} \psi^a \nabla_{ab} \psi^b-\frac{1}{16} \psi^d\psi_d\; \nabla^{ab}\psi^c\nabla_{ab}\psi_c =\nn \\
&=&1+\frac{1}{2} \psi^a \partial_{ab} \psi^b+\frac{1}{8}\psi^d\psi_d \left( \partial^{ab}\psi_b \partial_{ac}\psi^c+
\frac{1}{2} \partial^{ab}\psi^c \partial_{ab}\psi_c\right).
\eea

Let us stress, that such a simple form of the component action is achieved only in the rather specific basis, where the
bosonic $q$ and fermionic fields $\psi^a$ are the Goldstone fields of the nonlinear realization. Surely, this choice is not
unique and in  different bases  the explicit form of  action could drastically change. The most illustrative example is given by the action in \cite{AGIT}, where the on-shell component action for the supermembrane  has been constructed for the first time.

The detailed proof that the action \p{Action} is invariant with respect to both, broken and unbroken supersymmetries,
can be found in  Appendix B.
\setcounter{equation}{0}
\section{Supersymmetric D2 brane component action}
Due to the duality between scalar field and gauge field strength in $d=3$, the action for $D2$ brane can be easily constructed within the coset approach. The idea of the construction is similar to the purely bosonic case. The crucial step is to treat the first, bosonic component of $\lambda_{ab}$ as an independent component (i.e. to ignore the (a) part of Eqs.\p{ihA}).
Now, the generalized variant of the action \p{Action} reads
\be\label{d2act1}
S=\int d^3x \left[ 2- \det(E) -\det(E) \left( 1+ 2 \frac{\lambda^{ab}(\nabla_{ab}q +2 \lambda_{ab})}{1-2 \lambda^2}\right)\right].
\ee
All these summands have a description in terms of $\theta=0$ parts of the Cartan forms \p{cartan}. The first term is just a volume form constructed from ordinary differentials $dx^{ab}$. The second terms is a volume form constructed from
semi-covariant differentials $d{\hat x}{}^{ab}$
$$ d{\hat x}{}^{ab} = dx^{ab}+\frac{1}{2} \psi^a \partial_{cd} \psi^b dx^{cd}.$$
Finally, the last term in \p{d2act1} is a volume form constructed from $\theta=0$ component of the forms $\Omega_P^{ab}$ \p{cartan}
$$ d{\tilde x}^{ab} = d{\hat x}^{ab}+ \frac{2}{1-2\lambda^2} \lambda^{ab}\left( \nabla_{cd} q +2 \lambda_{cd}\right)d{\hat x}^{cd}. $$

Since the action \p{d2act1} depends only on $\lambda^{ab}$ and not on its derivatives, the $\lambda$-equation of motion
\be\label{D2eom1}
\nabla_{ab} q =- \frac{4\lambda_{ab}}{1+2 \lambda^2}
\ee
can be used to eliminate $\lambda^{ab}$ in favor of $\nabla_{ab} q$. Clearly, the equation \p{D2eom1} is just the
 $(a)$ part of the constraints \p{ihA}, we ignored while introducing the action \p{d2act1}. Plugging $\lambda$ expressed through $q$ back into \p{d2act1} gives us the action \p{Action}.

Alternatively, the equation of motion for $q$
\be\label{D2eom2}
\partial_{ab} \left[ \frac{ \det(E)\; \lambda^{cd}\; \left( E^{-1}\right)_{cd}{}^{ab}}{1-2\lambda^2}\right]=0
\ee
has the form of the $d=3$ Bianchi identity for the field strength $F^{ab}$
\be\label{D2FS}
F^{ab}\equiv \frac{ \det(E)\; \lambda^{cd}\; \left( E^{-1}\right)_{cd}{}^{ab}}{1-2\lambda^2} \qquad \Rightarrow \qquad
\partial_{ab} F^{ab}=0.
\ee
Substituting this into the action \p{d2act1} and integrating by parts, one may bring it to the supersymmetric $D2$-brane action
\be\label{BI}
S=\int d^3x \left[ 2 -\det(E)\left( 1+\sqrt{1+ 8 {\widetilde F}{}^2}\right)\right]
\ee
where
\be
{\widetilde F}_{ab} \equiv \frac{\left(E\right)_{ab}{}^{cd}\;F_{cd}}{\det(E)} = \frac{\lambda_{ab}}{1-2\lambda^2}.
\ee
Therefore,
\be\label{BII}
S=2 \int d^3 x \left[ 1- \det(E) \;\frac{1}{1-2 \lambda^2}\right].
\ee
Clearly, in the bosonic limit ${\widetilde F}_{ab}=F_{ab}$ and thus, the bosonic part of the \p{BI} is the standard Born-Infeld action for $D2$-brane, as it should be.
\setcounter{equation}{0}
\section{Conclusion}
In this Letter we proposed a method  to construct the on-shell component actions for  theories with $1/2$ partial
breaking of global supersymmetry within the nonlinear realization approach. In contrast with the standard superfield approach in which unbroken supersymmetry plays the leading role,  we have shifted the attention to the spontaneously broken supersymmetry. It turns out that in the theories in which half of supersymmetries are spontaneously broken, all physical fermions are just the fermions of the nonlinear realization. Moreover, the transformation properties of these fermions with respect to the broken supersymmetry are the same as in the famous Volkov-Akulov model. Just this fact completely fixed all possible appearances of the fermions in the component action: they can enter the action  through the determinant of the $d$-bein (to compensate the transformation of the volume form) and through  covariant derivatives, only. It is very important that in our parametrization of the coset the rest of the physical components, i.e. all the bosonic components, transform as  ``matter fields'' with respect to the broken supersymmetry. Clearly, in  such a situation the component action acquires the form of the Volkov-Akulov action for these ``matter fields''. The complete form of the action can be further fixed by two additional requirements: a) to reproduce the bosonic limit, which is explicitly known in many interesting cases, and b) to have a proper linearized form, which has to be invariant with respect to the linearized unbroken supersymmetry. We supply the general consideration by a detailed example of the component action of $N=1$ supermembrane in $D=4$ constructed within our  procedure. In this case we provide the readers with the  exact proof of the invariance of the component action with respect to both, broken and unbroken supersymmetries.

It should be clear that the extremely simple form of the component actions (at least in the explicit case of the supermembrane we considered in this Letter) is achieved due to the quite special choice of the physical components: all of them are  fields of the nonlinear realization.  This is in a dramatic contrast with the superfield approach, in which the main objects are the (super)fields of the linearly realized broken supersymmetry \cite{BG2, IK1, BIK1}. Of course, it is preferable to have
the superfield actions, but their very nice superspace forms become very complicated after passing to the components.
Moreover, in  such component actions it is a very nontrivial task to select some geometric objects and structures. In this respect, our construction looks as an alternative one, and the component form and on-shell character of our actions is the price we have to pay for their simplicity and clear geometric meaning.

It seems that the unique serious limitation of our construction is its validity for theories with $1/2$ breaking
of the global supersymmetries only. But even if it is so, the one explicit example given here is not enough to prove the efficiency of  our approach. Therefore, the evident
task is to apply the proposed method  to the cases where the superspace actions are known, having in mind to get more simple and understandable form of the components actions (e.g. supermembrane in $D=6$ \cite{BG2,BG1}, $N=1$ Born-Infeld action in $D=4$ \cite{BG2,RT}, superfield actions in AdS5 \cite{BIK2}, etc.). But the
most interesting cases are those  for which the components actions are still unknown (the action for partial breaking
of $N=1,D=10$ supersymmetry with the hypermultiplet as the Goldstone superfield \cite{BIK3}, etc.). Among the most
complicated and urgent task is the construction of $N=2$ Born-Infeld action within our scheme, using the nonlinear realization of \cite{N2BI}. Such an action has been recently constructed in \cite{RK} within a completely different approach. We are hoping that our variant of the action, when and if it will be constructed, would have a more simple structure.

Finally, it seems to be very important to make a careful comparison of the action we constructed in this Letter with known ones, constructed in \cite{AGIT} and those  which can be extracted from the explicit superfield action of \cite{IK1}.

\section*{Acknowledgements}
We wish to acknowledge discussions with Renata Kallosh and Nikolay Kozyrev.

S.K. and A.S. are grateful to the Laboratori Nazionali di Frascati for warm hospitality. This work was partially supported by RFBR grants~12-02-00517-a, 13-02-91330-NNIO-a and 13-02-90602 Apm-a, as well as by the ERC Advanced Grant no. 226455 \textit{``Supersymmetry, Quantum Gravity and Gauge Fields''}~(\textit{SUPER\-FIELDS}).

\setcounter{equation}{0}
\def\theequation{A.\arabic{equation}}
\section*{Appendix A: Superalgebra, coset space, transformations and Cartan forms}
In this Appendix we collected some formulas from the paper \cite{IK1} where the nonlinear realization
of $N=1, D=4$ Poincar\'{e} group in its coset over $d=3$ Lorentz group $SO(1,2)$ was constructed.

In $d=3$ notation the $N=1, d=D$  Poincar\'e superalgebra contains the following set of generators:
\be\label{1}
\mbox{ N=2, d=3 SUSY }\quad \propto \quad \left\{ Q_a, P_{ab}, S_a, Z, M_{ab}, K_{ab} \right\},
\ee
$a,b=1,2$ being the $d=3$ $SL(2,R)$
spinor indices \footnote{The indices are raised
and lowered as follows:
$V^{a}=\epsilon^{ab}V_b,\;V_{b}=\epsilon_{bc}V^c,\quad
\epsilon_{ab}\epsilon^{bc}=\delta_a^c\; .$}. Here, $P_{ab}$ and $Z$ are $D=4$ translation generators,  $Q_a$ and $S_a$ are the generators of super-translations, the generators $M_{ab}$ form $d=3$ Lorentz algebra $so(1,2)$, while the generators $K_{ab}$ belong to the coset $SO(1,3)/SO(1,2)$. The basic anticommutation relations read
\be
\left\{ Q_{a},Q_{b}\right\}=P_{ab}\; ,\quad
\left\{ Q_{a},S_{b}\right\} = \epsilon_{ab}Z\; , \quad
\left\{ S_{a},S_{b}\right\} = P_{ab} \;. \label{susy}
\ee

The coset element was
defined in \cite{IK1} as
\be\label{coset}
g=e^{x^{ab}P_{ab}}e^{\theta^{a}Q_{a}}e^{qZ}
  e^{\psi^aS_a}e^{\Lambda^{ab}K_{ab}} \;.
\ee
Here, $x^{ab}, \theta^a$ are $N=1, d=3$ superspace coordinates, while the remaining coset parameters are Goldstone superfields,
$\psi^a \equiv \psi^a(x,\theta),\;q \equiv q(x,\theta),\; \Lambda^{ab}
\equiv \Lambda^{ab}(x,\theta)$.

The transformation properties of the coordinates and superfields with respect
to all symmetries can be found by acting from the left on the coset element $g$ \p{coset} by the different elements of $N=1, D=4$ supergroup. They have the following explicit form:
\begin{itemize}
\item Unbroken supersymmetry $(g_0=\mbox{exp }(a^{ab}P_{ab}+
  \eta^{a}Q_{a} ))$
\be\label{susy1}
\delta x^{ab}=a^{ab}-\frac{1}{4}\eta^a\theta^b-\frac{1}{4}\eta^b\theta^a ,
\quad
\delta \theta^{a}=\eta^a\; .
\ee
\item Broken supersymmetry $(g_0=\mbox{exp }(\xi^{a}S_{a}))$
\be\label{susy2}
\delta x^{ab}= -\frac{1}{4}\xi^a\psi^b-\frac{1}{4}\xi^b\psi^a,\quad
\delta q=\xi^a\theta_a,\quad
\delta\psi^a=\xi^a \; .
\ee
\item $K$ transformations $(g_0=\mbox{exp }(r^{ab}K_{ab}))$
\bea
&&\delta x^{ab}= -2q r^{ab}-\frac{1}{2}\theta_c r^{ca}\psi^b-
   \frac{1}{2}\theta_c r^{cb}\psi^a +
     \frac{1}{2}\theta^a r^{bc}\psi_c+
     \frac{1}{2}\theta^b r^{ac}\psi_c\; , \nn\\
&&\delta \theta^{a} = -2r^{ab}\psi_b \; ,
\delta q = -4r_{ab}x^{ab} \; ,
\delta \psi^{ab}= 2r^{ab}\theta_b,\quad
   \delta \lambda^{ab}=r^{ab}-4 \lambda^{ac} r_{cd} \lambda^{db}\quad .
   \label{ktr}
\eea
\item Broken $Z$-translations $(g_0 = \mbox{exp}(cZ))$
\be \label{Ztr}
\delta q = c~.
\ee
\item The $d=3$ Lorentz group $SO(1,2)\sim SL(2,R)$ acts as rotations of
the spinor indices.
\end{itemize}
In \p{ktr} the coordinates of the stereographic parametrization of
the coset $SO(1,3)/SO(1,2)$ have been defined as
\be
\lambda^{ab}=
\frac{\tanh\left(\sqrt{2\Lambda^2}\right)}{\sqrt{2\Lambda^2}}\,
\Lambda^{ab}\; ,\quad
\tanh{}^2\left(\sqrt{2\Lambda^2}\right)\equiv 2 \lambda^2 \; ,\quad
\Lambda^2 \equiv \Lambda_{ab}\Lambda^{ab}~, \quad
\lambda^2 \equiv \lambda_{ab}\lambda^{ab}~.
\ee

The most important objects in the coset are the Cartan forms
$$g^{-1}d g =  \Omega_Q + \Omega_P + \Omega_Z + \Omega_S + \Omega_K +
\Omega_M .$$
In what follows we will need only the forms $\Omega_Q, \Omega_P, \Omega_Z$ and $\Omega_S$ which were constructed in \cite{IK1}
\bea
\Omega_Z & = & \frac{1+2\lambda^2}{1-2\lambda^2}\left[ d{\hat q}+
    \frac{4}{1+2\lambda^2}\lambda_{ab}d{\hat x}^{ab}\right]Z\; , \nn\\
\Omega_P &\equiv & \Omega_P^{ab}P_{ab} = \left[ d{\hat x}^{ab}+
\frac{2}{1-2\lambda^2}
\lambda^{ab}\left(
d{\hat q} + 2 \lambda_{cd}d{\hat x}^{cd}\right) \right] P_{ab} \; ,\nn \\
\Omega_Q &\equiv & \Omega_Q^aQ_a = \frac{1}{\sqrt{1-2\lambda^2}}
\left[ d\theta^{a}+
     2\lambda^{ab}d\psi_{b}\right]Q_{a} \; , \;
 \Omega_S  = \frac{1}{\sqrt{1-2\lambda^2}} \left[ d\psi^{a}-
     2\lambda^{ab}d\theta_{b}\right]S_{a} \;  .
\label{cartan} \\
d{\hat x}^{ab}  &\equiv &  dx^{ab}+\frac{1}{4}\theta^{a}d\theta^{b}
  +\frac{1}{4}\theta^{b}d\theta^{a}+
  \frac{1}{4}\psi^a d\psi^b +\frac{1}{4}\psi^{b}d\psi^{a} \; ,\quad
d{\hat q} \equiv  dq+\psi_{a}d\theta^{a} \; . \label{dx}
\eea
Note, that all Cartan forms, except for $\Omega_M$, transform homogeneously under all symmetries.

Having at hands the Cartan forms, one may construct the ``semi-covariant'' (covariant with respect to $d=3$ Lorentz,
unbroken and broken supersymmetries only) as
\be\label{cD}
d{\hat x}^{ab}\nabla_{ab} +d\theta^a \nabla_a = dx^{ab} \frac{\partial}{\partial x^{ab}} + d\theta^a \frac{\partial}{\partial \theta^a}.
\ee
Explicitly, they read \cite{IK1}
\be
\nabla_{ab} =  (E^{-1})^{cd}_{ab}\,\partial_{cd} \; , \quad
\nabla_a = D_a + \frac{1}{2}\psi^b D_a \psi^c \,\nabla_{bc} =
         D_a + \frac{1}{2}\psi^b \nabla_a \psi^c \,{\partial}_{bc}~, \label{nabla}
\ee
where
\bea
&&D_a=\frac{\partial}{\partial \theta^a}+
\frac{1}{2}\theta^b\partial_{ab}\; , \quad
\left\{ D_a, D_b \right\} =\partial_{ab} \; , \label{flatd} \\
&& E_{ab}^{cd}=\frac{1}{2}(\delta_a^c\delta_b^d+\delta_a^d\delta_b^c)+
  \frac{1}{4}(\psi^c\partial_{ab}\psi^d+ \psi^d\partial_{ab}\psi^c) \;,\label{E} \\
&& (E^{-1})^{cd}_{ab} = \frac{1}{2}(\delta_a^c\delta_b^d+\delta_a^d\delta_b^c)-
  \frac{1}{4}(\psi^c\nabla_{ab}\psi^d+ \psi^d\nabla_{ab}\psi^c) \;.
\eea
These derivatives obey the following algebra:
\bea
&& \left[ \nabla_{ab},\nabla_{cd} \right] =
-\nabla_{ab}\psi^m \nabla_{cd}\psi^n
           \nabla_{mn} \; , \qquad
\left[ \nabla_{ab},\nabla_{c} \right] =
\nabla_{ab}\psi^m\nabla_{c}\psi^n
           \nabla_{mn} \; , \nn \\
&& \left\{ \nabla_{a},\nabla_{b} \right\} =\nabla_{ab}+
        \nabla_{a}\psi^m\nabla_{b}\psi^n
           \nabla_{mn} \; .  \label{algebra}
\eea

To complete this rather technical Appendix, we will also define the  $d=3$ volume form  in a standard manner as
\be\label{volume}
d^3 x \equiv \epsilon_{ijk} dx^i \wedge dx^j \wedge dx^k \quad \Rightarrow \quad
dx^i \wedge dx^j \wedge dx^k = \frac{1}{6} \epsilon^{ijk} d^3 x.
\ee
The translation to the vectors is defined as
\be\label{44}
V^i \equiv \frac{i}{\sqrt{2}}\left(\sigma^i\right)_a{}^b\; V_b{}^a \quad \Rightarrow \quad V_a{}^b =-\frac{i}{\sqrt{2}} V^i \left( \sigma^i\right)_a{}^b,  \qquad V^{ab}V_{ab} = V^i V^i.
\ee
Here we are using the standard set of $\sigma^i$ matrices
\be\label{41}
\sigma^i \; \sigma^j = i \epsilon^{ijk} \sigma^k + \delta^{ij} E, \quad
\left( \sigma^i\right)_a{}^b \; \left(\sigma^i \right)_c{}^d = 2 \delta_a{}^d \delta_c{}^b - \delta_a{}^b \delta_c{}^d,
\ee
were $\epsilon^{ijk}$ obeys relations
\be\label{42}
\epsilon^{ijk}\epsilon^{imn} =\delta^{jm}\delta^{kn} -\delta^{jn}\delta^{km},\quad \epsilon^{ijk}\epsilon^{ijn}=2 \delta^{kn}, \quad \epsilon^{ijk}\epsilon^{ijk}=6.
\ee
\setcounter{equation}{0}
\def\theequation{B.\arabic{equation}}
\section*{Appendix B}
In this Appendix we will prove the invariance of the supermembrane action \p{Action} under broken and unbroken supersymmetries. The proof for the broken supersymmetry is the easiest one and we will start with this invariance.
\subsection*{Broken supersymmetry}
Under spontaneously broken $S^a$ supersymmetry our coordinates and the physical components transform as in \p{susy2}, i.e.
\be\label{Bsusy2}
\delta x^{ab}= -\frac{1}{4}\xi^a\psi^b-\frac{1}{4}\xi^b\psi^a,\quad
\delta q=0,\quad \delta\psi^a=\xi^a \; .
\ee
One may immediately check that the $\theta=0$ part of the covariant differential $d{\hat x}^{ab}$, defined in \p{dx}
\be\label{Bdx}
d{\hat x}^{ab}  =   dx^{ab}+
  \frac{1}{4}\psi^a d\psi^b +\frac{1}{4}\psi^{b}d\psi^{a}
\ee
is invariant under the transformations \p{Bsusy2}. Therefore, the covariant derivatives $\nabla_{ab}$ \p{nabla} are also invariant under broken supersymmetry transformations. Now, for the active form of the transformations
($\delta \phi =\phi'(x)-\phi(x)$) we have
\bea\label{Trsusy2}
&& \delta_S \nabla^{ab} q = \frac{1}{2} \xi^c \psi^d \partial_{cd} \nabla^{ab}q \quad \Rightarrow\quad
\delta_S{\cal F}(\nabla q \cdot \nabla q) =\frac{1}{2} \xi^a \psi^b \partial_{ab} {\cal F}, \nn \\
&& \delta_S \psi^a = \xi^a+\frac{1}{2} \xi^c \psi^d \partial_{cd} \psi^a, \qquad
\delta_S \nabla^{ab} \psi^c = \frac{1}{2} \xi^d \psi^e \partial_{de} \nabla^{ab}\psi^c,
\eea
and, therefore,
\be\label{trdetE}
\delta_S \det(E) = \frac{1}{2} \xi^a \nabla_{ab}\psi^b - \frac{1}{8} \xi^d \psi_d \nabla^{ab}\psi^c\nabla_{ab}\psi_c +
\frac{1}{2} \xi^c \psi^d \partial_{cd} \det(E).
\ee
Thus, the integrand in the action \p{cact1} transforms as follows:
\bea\label{tract1}
\delta_S\left( \det(E) {\cal F}\right)&=& \left( \frac{1}{2} \xi^a \nabla_{ab}\psi^b - \frac{1}{8} \xi^d \psi_d \nabla^{ab}\psi^c\nabla_{ab}\psi_c\right) {\cal F}+\frac{1}{2} \xi^c \psi^d \partial_{cd}\left[ \det(E){\cal F}\right] = \nn \\
&& \left( \frac{1}{2} \xi^a \nabla_{ab}\psi^b - \frac{1}{8} \xi^d \psi_d \nabla^{ab}\psi^c\nabla_{ab}\psi_c -
\frac{1}{2} \xi^c \partial_{cd}\psi^d \det(E)\right) {\cal F}.
\eea
It is a matter of direct calculations to check that the expression in the parentheses in \p{tract1} is zero. Thus, the action \p{cact1}, as well as the action \p{Action}, are indeed invariant under spontaneously broken supersymmetry.
\subsection*{Unbroken supersymmetry}
It is funny, but in contrast with the superfield approach in which unbroken supersymmetry is manifest, to prove the invariance of the component action \p{Action} under unbroken supersymmetry is a rather complicated task.

Under unbroken $(Q^a)$ supersymmetry the covariant derivatives $\nabla_{ab}, \nabla_a$ \p{nabla} are invariant by construction. Therefore, the objects $\nabla_{ab} \psi_c, \nabla_{ab} q$ are the superfields with the standard transformation
properties:
\bea
&& \delta_Q \psi^a \equiv -\eta^b D_b \psi^a = 2 \eta^b\left( \lambda_b{}^a -\frac{1}{2} \psi^m \lambda_b{}^n\partial_{mn}\psi^a\right), \label{Qpsi} \\
&& \delta_Q \nabla_{ab}\psi_c \equiv -\eta^d D_d \nabla_{ab}\psi_c=-\eta^d\left( 2 \nabla_{ab}\psi^m \lambda_d{}^n \nabla_{mn}\psi_c-2\nabla_{ab}\lambda_{dc}+
\psi^m\lambda_d{}^n \partial_{mn} \nabla_{ab}\psi_c\right), \label{Qnablapsi} \\
&& \delta_Q \nabla_{ab}q \equiv -\eta^c D_c \nabla_{ab}q=-\eta^c\left( \frac{1-2\lambda^2}{1+2\lambda^2}\nabla_{ab}\psi_c+ \psi^m\lambda_c{}^n\partial_{mn} \nabla_{ab} q \right). \label{Qq}
\eea
Therefore,
\bea\label{QdetE}
\delta_Q \det(E)&=& \eta^c\lambda_c{}^a \nabla_{ab}\psi^b -\eta^c \nabla_{ab}\lambda^b{}_c \psi^a +
\eta^c\lambda_c{}^n \psi^a \nabla_{ab}\psi^m \nabla_{mn}\psi^b-\frac{1}{4}\eta^b\lambda_b{}^a \psi_a \nabla^{mn}\psi^k \nabla_{mn}\psi_k-\nn \\
&& \frac{1}{8} \psi^2 \eta^d \lambda_d{}^b \nabla_{bc}\psi^c \nabla^{mn}\psi^k \nabla_{mn}\psi_k+
\frac{1}{4} \psi^2 \eta^d \nabla_{ab}\lambda_{dc} \nabla^{ab} \psi^c-\eta^c\lambda_c{}^n\psi^m \partial_{mn} \det(E),
\eea
and
\be\label{QF}
\delta_Q {\cal F}= -2 \frac{1-2 \lambda^2}{1+2\lambda^2} \eta^c \nabla_{ab}\psi_c \nabla^{ab} q {\cal F}'-
\eta^c\lambda_c{}^n \psi^m \partial_{mn} {\cal F}.
\ee
The ${\cal F}'$ in \p{QF} denotes the derivative ${\cal F}$ over its argument (i.e. over $\nabla q \cdot \nabla q$ in our
case).

Combining these expressions we will get the following variation of the integrand of our action \p{Action}:
\be\label{add1}
\delta_Q {\cal L} \equiv \delta_Q\left( \det(E) {\cal F}\right) = \delta_Q \det(E) {\cal F} + \det(E)\delta_Q{\cal F}.
\ee
In \p{add1} the last terms from $\delta_Q \det(E)$ \p{QdetE} and $\delta_Q{\cal F}$ \p{QF} combine together to produce
$$ -\eta^a \lambda_{a}{}^b \psi^c \partial_{bc} \left[ \det(E){\cal F}\right].$$
Therefore, after integration by parts in this term we will get
\bea\label{add2}
\delta_Q {\cal L}&=& \left( \eta^c\lambda_c{}^a \nabla_{ab}\psi^b -\eta^c \nabla_{ab}\lambda^b{}_c \psi^a +
\eta^c\lambda_c{}^n \psi^a \nabla_{ab}\psi^m \nabla_{mn}\psi^b-\frac{1}{4}\eta^b\lambda_b{}^a \psi_a \nabla^{mn}\psi^k \nabla_{mn}\psi_k-\right.\nn \\
&&\left. \frac{1}{8} \psi^2 \eta^d \lambda_d{}^b \nabla_{bc}\psi^c \nabla^{mn}\psi^k \nabla_{mn}\psi_k+
\frac{1}{4} \psi^2 \eta^d \nabla_{ab}\lambda_{dc} \nabla^{ab} \psi^c\right) {\cal F}- \\
&& 2 \frac{1-2 \lambda^2}{1+2\lambda^2} \eta^c \nabla_{ab}\psi_c \nabla^{ab} q {\cal F}' \det(E)+
\eta^c \partial_{mn} \lambda_c{}^n \psi^m {\cal F} \det(E)+\eta^c\lambda_c{}^n \partial_{mn} \psi^m  {\cal F} \det(E).\nn
\eea
Now, one may check that three terms with the derivatives of $\lambda_{ab}$ (i.e., the second terms in each of all three lines of \p{add2}) just canceled.

The next step is to substitute into \p{add2} the explicit expressions for $\lambda_{ab}$ \p{ihA} and for ${\cal F}$ \p{Action}
\be\label{add3}
\lambda_{ab}=\frac{-\frac{1}{2} \nabla_{ab} q}{1+\sqrt{1-\frac{1}{2} \nabla q \cdot \nabla q}}, \qquad
{\cal F}= 1+\sqrt{1-\frac{1}{2} \nabla q \cdot \nabla q}.
\ee
If we note that
\be\label{add4}
\lambda_{ab}=\frac{-\frac{1}{2} \nabla_{ab} q}{\cal F} \qquad \mbox{and} \qquad
\frac{1-2 \lambda^2}{1+2\lambda^2} =-\frac{1}{4\; {\cal F}'},
\ee
it will be not so strange that after substitution of \p{add3} into \p{add2}, the variation $\delta_Q {\cal L}$ will not contain any square roots. So, it will  read
\bea\label{add5}
\delta_Q {\cal L}&=& -\frac{1}{2}\eta^c\nabla_c{}^a q \nabla_{ab}\psi^b -
\frac{1}{2} \eta^c \nabla_{c}{}^n q \psi^a \nabla_{ab}\psi^m \nabla_{mn}\psi^b+
\frac{1}{8}\eta^b \nabla_b{}^a q \psi_a \nabla^{cd}\psi^e \nabla_{cd}\psi_e+ \nn \\
&& \frac{1}{16} \psi^2 \eta^a \nabla_a{}^b q \nabla_{bc}\psi^c \nabla^{de}\psi^f \nabla_{de}\psi_f+
\frac{1}{2} \eta^c \nabla_{ab} \psi_c \nabla^{ab} q \det(E)- \nn \\
&& \frac{1}{2} \eta^a \nabla_a{}^b q\; \partial_{bc}\psi^c \det(E).
\eea
Substituting now the expression for $\partial_{bc}\psi^c \det(E)$ from \p{tract1} and slightly rearranging the terms,
we obtain
\bea\label{add6}
\delta_Q {\cal L}&=& -\eta^c\nabla_c{}^a q \nabla_{ab}\;\psi^b -
\frac{1}{4} \eta^a \nabla_{a}{}^b q \;\psi_b \nabla^{cd}\psi_d \nabla_{ce}\psi^e+
 \frac{1}{16} \psi^2 \eta^a \nabla_a{}^b q \nabla_{bc}\psi^c \nabla^{de}\psi^f \nabla_{de}\psi_f+ \nn \\
&&  \frac{1}{2} \eta^c \nabla_{ab} \psi_c \nabla^{ab} q \det(E).
\eea
Finally, combining the terms in the first line together, we will get the following simple form of the variation of
the integrand:
\be\label{varL}
\delta_Q {\cal L} = -\eta^c \left( \nabla_c{}^a q\nabla_{ab} \psi^b-
\frac{1}{2} \nabla_c{}^{n} q \nabla_{ab} \psi_c\right) \det(E).
\ee
Unfortunately,  further simplifications are not possible. The simplest way to be sure that $\delta_Q {\cal L}$ \p{varL}
gives zero after integration over $d^3x$ is to find the ``equation of motion'' for $q$ which follows from the
``Lagrangian''  \p{varL}
\be\label{add7}
\frac{\delta}{\delta q} \int d^3x \;\delta_Q {\cal L} =0.
\ee
Clearly, expression \p{add7} has to be identically equal to zero if our action is invariant under unbroken supersymmetry.
After quite lengthly and tedious, but straightforward calculations, one may show that this is indeed so.

Thus, our action \p{Action} is invariant with respect both, broken and unbroken supersymmetries.


\begin{thebibliography}{99}
\bibitem{dbrane} M.~Aganagic, C.~Popescu, J.H.~Schwarz,\\
{\it Gauge-Invariant and Gauge-Fixed D-Bane Actions},\\
Nucl.Phys. {\bf B495} (1997) 99, {\tt arXiv:hep-th/9612080}.
\bibitem{NR}
S.R.~Coleman, J.~Wess, B.~Zumino,\\
{\it Structure of phenomenological Lagrangians. 1},\\
Phys.Rev. 177 (1969) 2239,\\
{\it Structure of phenomenological Lagrangians. 2},\\
Phys.Rev. 177 (1969) 2247.
\bibitem{NR1} D.V.~Volkov,\\
{\it Phenomenological Lagrangians},\\
Sov.J.Part.Nucl. {\bf 4}(1973) 3;\\
V.I.~Ogievetsky,\\
``Nonlinear realizations of internal and space-time symmetries'',\\
In Proceedings of the Xth Winter School of Theoretical Physics in Karpacz, Vol.1, p.117, 1974.
\bibitem{ih} E.A. Ivanov, V.I. Ogievetsky, \\
{\it The Inverse Higgs Phenomenon in Nonlinear Realizations},\\
Teor. Mat. Fiz. {\bf 25} (1975) 164.
\bibitem{VA} D.V.~Volkov and V.P.~Akulov,\\
{\it Possible universal neutrino interaction},\\
JETP Lett. {\bf 16}(1972 438; \\
{\it Is the neutrino a Goldstone particle?}\\
Phys. Lett. {\bf B46}(1973) 109.
\bibitem{CNV} T.E.~Clark, M.~Nitta,T.~ter~Veldhuis,\\
{\it Brane dynamics from nonlinear realizations},\\
Phys.Rev. {\bf D67} (2003) 085026.
\bibitem{Iv} S.~Bellucci, E.~Ivanov, S.~Krivonos,\\
{\it Superbranes and Super Born-Infeld Theories from Nonlinear Realizations},\\
 Nucl.Phys.Proc.Suppl. {\bf 102} (2001) 26; {\tt   arXiv:hep-th/0103136}.
\bibitem{Dima} D.~Sorokin,\\
{\it Superbranes and Superembeddings},\\
Phys.Rept. {\bf 329} (2000) 1; {\tt  arXiv:hep-th/9906142}.
\bibitem{BG2}J.~Bagger, A.~Galperin, \\
{\it New Goldstone multiplet for partially broken supersymmetry},\\
Phys.Rev. {\bf D55} (1997) 1091; {\tt  arXiv:hep-th/9608177}.
\bibitem{RT}M.~Rocek, A.A.~Tseytlin, \\
{\it Partial breaking of global D=4 supersymmetry, constrained superfields, and 3-brane actions},\\
Phys.Rev. {\bf D59} (1999) 106001; {\tt  arXiv:hep-th/9811232}.
\bibitem{IK1} E.~Ivanov, S.~Krivonos,\\
{\it N=1, D=4 supermembrane in the coset approach},\\
Phys.Lett {\bf B453} (1999) 237, {\tt arXiv:hep-th/9901003}.
\bibitem{BIK1} S.~Bellucci, E.~Ivanov, S.~Krivonos, \\
{\it Towards the complete N=2 superfield Born-Infeld action with partially broken N=4 supersymmetry},\\
Phys.Rev. {\bf D64} (2001) 025014; {\tt  arXiv:hep-th/0101195}.
\bibitem{RK}E.~Bergshoeff, F.~Coomans, R.~Kallosh, C.S.~Shahbazi, A.~Van~Proeyen,\\
{\it Dirac-Born-Infeld-Volkov-Akulov and Deformation of Supersymmetry},\\
arXiv:1303.566[hep-th].
\bibitem{AGIT} A.~Ach\'{u}carro, J.~Gauntlett, K.~Itoh and P.K.~Townsend,\\
{\it World-Volume Supersymmetry from Spacetime Supersymmetry of the Four-Dimensional Supermembrane},\\
Nuclear Physics {\bf B314} (1989) 129.
\bibitem{BG1} J.~Bagger, A.~Galperin,\\
{\it Matter couplings in partially broken extended supersymmetry},\\
Phys.Lett. {\bf B336} (1994) 25-31; {\tt arXiv:hep-th/9406217}.
\bibitem{BIK2} S.~Bellucci, E.~Ivanov, S.~Krivonos,\\
{\it Goldstone Superfield Actions in AdS5 backgrounds},\\
Nucl.Phys. {\bf B672} (2003) 123; {\tt arXiv:hep-th/0212295}.
\bibitem{BIK3} S.~Bellucci, E.~Ivanov, S.~Krivonos,\\
{\it Partial breaking of N=1, D=10 supersymmetry}, \\
Phys.Lett. {\bf B460} (1999) 348; {\tt arXiv:hep-th/9811244},\\
{\it Partial breaking N=4 to N=2: hypermultiplet as a Goldstone superfield},\\
Fortsch.Phys. {\bf 48} (2000)19; {\tt arXiv:hep-th/9809190}.
\bibitem{N2BI} S.~Bellucci, E.~Ivanov, S.~Krivonos,\\
{\it N=2 and N=4 supersymmetric Born-Infeld theories from nonlinear realizations},\\
Phys.Lett. {\bf B502} (2001) 279; {\tt arXiv:hep-th/0012236}.
\end{thebibliography}
\end{document}